# Rapid infrared imaging for rhombohedral graphene

Zuo Feng[1,2,†], Wenxuan Wang[2,†], Yilong You[1,2,†], Yifei Chen[3], Kenji Watanabe[4], Takashi Taniguchi[5], Chang Liu[1,2]*, Kaihui Liu[1,2]* and Xiaobo Lu[2,6]*

[1]State Key Laboratory for Mesoscopic Physics, Frontiers Science Centre for Nano-optoelectronics, School of Physics, Peking University, 100871, Beijing, China
[2]International Center for Quantum Materials, School of Physics, Peking University, Beijing 100871, China
[3]International School of Beijing, Beijing, 101318
[4]Research Center for Electronic and Optical Materials, National Institute of Material Sciences, 1-1 Namiki, Tsukuba 305-0044, Japan
[5]Research Center for Materials Nanoarchitectonics, National Institute of Material Sciences, 1-1 Namiki, Tsukuba 305-0044, Japan
[6]Collaborative Innovation Center of Quantum Matter, Beijing 100871, China

[†]Z.F., W.W., and Y.L. contributed equally to this work.
*E-mail: xiaobolu@pku.edu.cn; liu.chang@pku.edu.cn; khliu@pku.edu.cn

**The extrinsic stacking sequence based on intrinsic crystal symmetry in multilayer two-dimensional materials plays a significant role in determining their electronic and optical properties. Compared with Bernal-stacked (ABA) multilayer graphene, rhombohedral (ABC) multilayer graphene hosts stronger electron-electron interaction due to its unique dispersion at low-energy excitations and has been utiliazed as a unique platform to explore strongly correlated physics. However, discerning the stacking sequence has always been a quite time-consuming process by scanning mapping methods. Here, we report a rapid recognition method for ABC-stacked graphene with high accuracy by infrared imaging based on the distinct optical responses at infrared range. The optical contrast of the image between ABC and ABA stacked graphene is strikingly clear, and the discernibility is comparable to traditional optical Raman microscopy but with higher consistency and throughput. We further demonstrate that the infrared imaging technique can be integrated with dry transfer techniques commonly used in the community. This rapid and convenient infrared imaging technique will significantly improve the sorting efficiency for differently stacked multilayer graphene, thereby accelerating the exploration of the novel emergent correlated phenomena in ABC stacked graphene.**

Rhombohedral (ABC) stacked multilayer graphene has been a highly tunable platform to explore strongly correlated and topological phenomena. At low-energy excitation, the dispersion of ABC stacked multilayer graphene follows the relation of $E=Ap^N$ for single-particle level, where $E$ is the kinetic energy, $A$ is a constant, $p$ is the momentum and $N$ is the number of layers. The unique band dispersion leads to strong correlation between electrons dominating over their kinetic energy at low Fermi energy, particularly for the layer number $N \geq 3$ [1-8]. In recent years, a plethora of novel correlated phenomena have come to light in ABC stacked graphene systems, including Mott insulators[5], layer-polarized antiferromagnetic insulators [9-11], superconductivity [12, 13], Stoner ferromagnetism [14], orbital multiferroicity[15], integer Chern insulators [16-18] and fractional Chern insulators [19, 20] based on a small number of devices reported by several laboratories. So far, the fabrication of standard ABC graphene devices compatible with transport measurement is still challenging due to two reasons: (i) the energetically metastable ABC stacked sequence will easily relax into Bernal stacked (ABA) sequence with lower energy during the fabrication processes; (ii) identifying ABC stacked region of graphene flakes is time-consuming and involves scanning techniques, such as scanning near-field optical microscopy [21-23], scanning Raman spectroscopy [24-26], scanning Kelvin probe microscopy [27], second harmonic generation [28, 29] and angle-resolved photoemission spectroscopy [30]. Developing a rapid detection method for discerning the graphene stacking sequence, both before and after the assembly process of the heterostructures, would significantly facilitate device fabrication.

In this study, we have reported a real-time imaging method for rapidly distinguishing the stacking sequence of multilayer graphene in the infrared light range. The infrared imaging technique provides high-quality images of differently stacked domains (ABC or ABA stacking) with a resolution better than 1 μm, consistent with results obtained from Raman spectroscopy mapping. Moreover, the ABC and ABA stacking sequence can also be distinguished when graphene flakes are encapsulated with hBN flakes, allowing for convenient monitoring stacking sequence changes during assembly process. By employing the infrared imaging method, four devices of rhombohedral graphene with different layers were successfully fabricated, which exhibit characteristic phenomena of ABC stacked graphene.

In ABC stacked graphene, the second layer atoms sit in the center of the hexagonal lattice of the first layer atoms, and the third layer atoms continue to shift in the same direction. In contrast, in ABA stacked graphene, the third layer aligns directly above the first layer (Fig. 1a and b). The different stacking sequences result in radically different band structures for multilayer graphene [31, 32]. For tetralayer graphene, ABA stacked graphene has two hyperbolic bands near the K point and two split-off bands while ABC graphene has one hyperbolic band and three split-off bands (Fig. 1c and d). Details of the tight-binding calculation are shown in supplementary information (Supplementary Note 1). The optical conductivity of a material arises from two distinct types of absorption mechanisms: the intraband absorption and interband absorption. Interband transition describes the process in which an electron absorbs a photon and jumps from the valence band to the conduction band directly. Intraband absorption requires the scattering with phonons to satisfy the momentum conservations. In graphene, optical conductivity is primarily attributed to intraband transitions in the far-infrared region. In contrast, for visible and near-infrared wavelengths, optical conductivity is predominantly determined by interband transitions [33-37]. In Fig. 1e and f, we plot the interband transition distribution for photons with energy of 0.78 eV (corresponding to wavelength of 1600 nm) for ABC tetralayer graphene with two different stacking sequences. For ABC stacked graphene, the transition from the V1 band to the C1 band accounts for ~40% of the optical conductivity contribution. The transition from the V2 band to the C2 band accounts for ~25% of the optical conductivity contribution. For ABA stacked graphene, the transition from the V1 band to the C1 band and from the V2 band to the C2 band both contribute approximately 40% of the optical conductivity. Considering all the optical transitions, ABC stacked tetralayer graphene exhibits lower optical conductivity compared to ABA stacked graphene in the infrared range (Fig. 1g). The optical conductivity of ABC stacked graphene decreases with increasing wavelength, while the optical conductivity of ABA stacked graphene increases. Specifically, in the range of 1200 nm to 1600 nm, the conductivity of ABA stacked graphene is consistently higher than that of ABC stacked graphene.

We have further experimentally investigated the infrared properties of ABA and ABC stacked graphene (including trilayer, tetralayer and decalayer graphene) with differential reflectance spectroscopy covering wavelengths from 1200 nm to 1600 nm. In our experiment, the multilayer graphene was exfoliated onto a highly doped silicon substrate with 285 nm oxide layer which is widely used in the community. The stacking sequence was initially confirmed with Raman spectrum. The differential reflectance $\Delta R/R$ is defined as $\Delta R/R=(R_{gr}-R_{sub})/R_{sub}$, where $R_{gr}$ is the reflectance of graphene on substrate, $R_{sub}$ is the reflectance of bare substrate. Fig. 2a shows the calculated differential reflectance $\Delta R/R$ considering the multi-reflection process for a monolayer graphene. We use a four-layer model composed of air ($n_0$), graphene ($n_{gr}$), silicon oxide ($n_2$) and silicon substrate ($n_3$). More details are shown in supplementary information. Theoretically, $R_{gr}=|(r_1+r_2e^{-2i\beta_1}+r_3e^{-2i(\beta_1+\beta_2)}+r_1r_2r_3e^{-2i\beta_2})/(1+r_1r_2e^{-2i\beta_1}+r_1r_3e^{-2i(\beta_1+\beta_2)}+r_2r_3e^{-2i\beta_2})|^2$, where $r_1=(n_0-n_{gr})/(n_0+n_{gr})$, $r_2=(n_{gr}-n_2)/(n_{gr}+n_2)$, $r_3=(n_2-n_3)/(n_2+n_3)$ are the reflection coefficients between the interface, respectively. $\beta_1=2\pi n_{gr}d_1/\lambda$, $\beta_2=2\pi n_2d_2/\lambda$ are the phase changes due to the different optical path. The refractive index of graphene ($n_{gr}$) sensitively depends on the optical conductivity by the relation $\sigma = i\omega(n_{gr}^2 - \varepsilon_0)$, where $\omega$ is the frequency of incident light, $\varepsilon_0$ is the dielectric constant of vacuum. For multilayer graphene, different values of optical conductivity $\sigma$ originating from different stacking sequence (Fig. 1g) can lead to distinct different values of $\Delta R/R$ as shown in Fig. 2b-d.

Based on the distinct differential reflectance results, we have further designed a real-time infrared imaging system to distinguish different stacking sequences of multilayer graphene. As shown in Fig.3a, the infrared light with effective wavelength $\lambda>900$nm achieved by a long-wave-pass filter is used for illumination. The reflected signal is collected through an objective with a magnification value of 100 and a numerical aperture value of 0.85. The signal is further recorded with an InGaAs CCD with a detection range from 900 nm to 1700 nm.

Representative graphene samples, including trilayer, heptalayer and hBN encapsulated tetralayer graphene, were characterized with the infrared imaging system. Fig. 3b and c show the optical image taken from traditional optical microscopy which can only reveal uniform optical contrast. Remarkably, a sharp contrast can be revealed with the infrared imaging system due to the different optical reflectance of ABC stacked and ABA stacked graphene in the infrared range. As shown in Fig. 3e and f, the brighter areas correspond to ABC stacked regions (indicated by the blue dashed lines), while the darker areas correspond to ABA stacked regions. The 2D mode of Raman spectra in ABC stacked graphene exhibits a more asymmetric feature with a higher shoulder at lower frequencies compared with that in ABA stacked graphene (Fig. 3k and i) [24]. The mapping of full width at half maximum (FWHM) of the 2D mode in Raman spectra haven been used as an independent way to distinguish the regions with different stacking sequence (Fig. 3h and i). The infrared imaging results are highly consistent with Raman spectroscopy mapping results. To verify the universality of the imaging technique, multilayer graphene samples with layer number ranging from 3-10 have been imaged with the infrared imaging system. The contrast differences between the two stacking sequences ranged from 10% to 40%, making them easily identifiable. We have further demonstrated the infrared imaging system as a reliable and versatile technique to distinguish graphene stacking sequences when encapsulated with hBN flakes. Fig. 3d shows the optical image of tetralayer graphene encapsulated with hBN and gated with a bottom graphite gate. The insert of Fig. 3d shows the cross-section of the stack. To increase the yield of retaining ABC stacking regions, the graphene flake was cut into different small square areas. Fig. 3g shows the infrared imaging results of the sample. The region indicated with blue dashed box in Fig. 3g exhibits brighter contrast originating from ABC stacking sequence, which is in good agreement with Raman spectroscopy measurement shown in Fig. 3j and m.

With the assist of the rapid infrared imaging technique, the pure ABC stacking regions were accurately selected for further device fabricating. Four devices of ABC stacked graphene, with thickness ranging from three to six layers, were fabricated. Fig. 4a schematically shows the cross-section of the devices. The carrier density $n = C_{tg}V_{tg}+C_{bg}V_{bg}$ and displacement field $D = (C_{tg}V_{tg} - C_{bg}V_{bg})/2\varepsilon_0$ can be independently tuned by top gate voltage ($V_{tg}$) and bottom gate voltage ($V_{bg}$) applied through the two graphite flaks. Here $C_{tg}$ ($C_{bg}$) is the capacitance between top (bottom) gate and graphene, $\varepsilon_0$ is the vacuum dielectric constant and $e$ is the elementary charge. The optical images of these devices are shown in Fig. 4b-e, respectively. Fig. 4f-i shows the four-terminal resistance as a function of carrier density $n$ and displacement field $D$ measured from the four devices at a temperature of T = 1.5K. The ABC trilayer, tetralayer graphene and pentalayer graphene were designed to aligned to the top hBN to form moiré superlattices, showing qualitatively similar correlated states at some integer moiré fillings with previous studies[5, 13, 14, 19]. The hexalayer device shows an insulating state at $D = 0$ and $n = 0$, which is similar with tetralayer graphene and pentalayer graphene and has been attributed to a layer-antiferromagnetic state[9, 11].

In summary, we have introduced a rapid imaging technique for discerning ABC stacked graphene. The ABC stacked regions exhibit a brighter contrast under infrared light compared to ABA stacked regions on $SiO_2$/Si substrate. The stacking sequence of graphene encapsulated by hBN can also be distinguished during the whole fabrication process. With the help of this technique, four ABC devices with high quality were successfully fabricated. We believe that this technique will significantly enhance the throughput of device fabrication and advance the exploration of strongly correlated physics emerged in ABC stacked multilayer graphene.

Acknowledges support from the National Key R&D Program (Grant nos. 2022YFA1403500/02) and the National Natural Science Foundation of China (12141401, 52025023, 11888101, T2188101).

X.L. and K.L. supervised the projects. Z.F. and W.W. fabricated the devices and performed the measurement. Y.Y, Y.C and C.L. performed the theoretical modeling; K.W., T.T contributed materials; Z.F. and X.L. wrote the paper with input from all authors.

References
1. Koshino, M., *Interlayer screening effect in graphene multilayers withABAandABCstacking.* Physical Review B, 2010. **81**(12).
2. Mak, K.F., J. Shan, and T.F. Heinz, *Electronic structure of few-layer graphene: experimental demonstration of strong dependence on stacking sequence.* Phys Rev Lett, 2010. **104**(17): p. 176404.
3. Bao, W., et al., *Stacking-dependent band gap and quantum transport in trilayer graphene.* Nature Physics, 2011. **7**(12): p. 948-952.
4. Lee, Y., et al., *Competition between spontaneous symmetry breaking and single-particle gaps in trilayer graphene.* Nature Communications, 2014. **5**(1).
5. Chen, G., et al., *Evidence of a gate-tunable Mott insulator in a trilayer graphene moiré superlattice.* Nature Physics, 2019. **15**(3): p. 237-241.
6. Lui, C.H., et al., *Observation of an electrically tunable band gap in trilayer graphene.* Nature Physics, 2011. **7**(12): p. 944-947.
7. Aoki, M. and H. Amawashi, *Dependence of band structures on stacking and field in layered graphene.* Solid State Communications, 2007. **142**(3): p. 123-127.
8. Winterer, F., et al., *Ferroelectric and spontaneous quantum Hall states in intrinsic rhombohedral trilayer graphene.* Nature Physics, 2024. **20**(3): p. 422-427.
9. Liu, K., et al., *Spontaneous broken-symmetry insulator and metals in tetralayer rhombohedral graphene.* Nature Nanotechnology, 2023.
10. Zhou, W., et al., *Layer-polarized ferromagnetism in rhombohedral multilayer graphene.* Nat Commun, 2024. **15**(1): p. 2597.
11. Han, T., et al., *Correlated insulator and Chern insulators in pentalayer rhombohedral-stacked graphene.* Nature Nanotechnology, 2023. **19**(2): p. 181-187.
12. Zhou, H., et al., *Superconductivity in rhombohedral trilayer graphene.* Nature, 2021. **598**(7881): p. 434-438.
13. Chen, G., et al., *Signatures of tunable superconductivity in a trilayer graphene moire superlattice.* Nature, 2019. **572**(7768): p. 215-219.
14. Zhou, H., et al., *Half- and quarter-metals in rhombohedral trilayer graphene.* Nature, 2021. **598**(7881): p. 429-433.
15. Han, T., et al., *Orbital multiferroicity in pentalayer rhombohedral graphene.* Nature, 2023. **623**(7985): p. 41-47.

16. Sha, Y., et al., *Observation of a Chern insulator in crystalline ABCA-tetralayer graphene with spin-orbit coupling.* Science, 2024. **384**(6694): p. 414-419.

17. Han, T., et al., *Large quantum anomalous Hall effect in spin-orbit proximitized rhombohedral graphene.* Science, 2024. **384**(6696): p. 647-651.

18. Chen, G., et al., *Tunable correlated Chern insulator and ferromagnetism in a moiré superlattice.* Nature, 2020. **579**(7797): p. 56-61.

19. Lu, Z., et al., *Fractional quantum anomalous Hall effect in multilayer graphene.* Nature, 2024. **626**(8000): p. 759-764.

20. Xie, J., et al. *Even- and Odd-denominator Fractional Quantum Anomalous Hall Effect in Graphene Moire Superlattices.* 2024. arXiv:2405.16944 DOI: 10.48550/arXiv.2405.16944.

21. Yin, L.-J., H. Jiang, J.-B. Qiao, and L. He, *Direct imaging of topological edge states at a bilayer graphene domain wall.* Nature Communications, 2016. **7**(1).

22. Jiang, L., et al., *Manipulation of domain-wall solitons in bi- and trilayer graphene.* Nature Nanotechnology, 2018. **13**(3): p. 204-208.

23. Li, H., et al., *Global Control of Stacking-Order Phase Transition by Doping and Electric Field in Few-Layer Graphene.* Nano Lett, 2020. **20**(5): p. 3106-3112.

24. Lui, C.H., et al., *Imaging stacking order in few-layer graphene.* Nano Lett, 2011. **11**(1): p. 164-9.

25. Nguyen, T.A., J.-U. Lee, D. Yoon, and H. Cheong, *Excitation Energy Dependent Raman Signatures of ABA- and ABC-stacked Few-layer Graphene.* Scientific Reports, 2014. **4**(1).

26. Cong, C., et al., *Raman Characterization of ABA- and ABC-Stacked Trilayer Graphene.* ACS Nano, 2011. **5**(11): p. 8760-8768.

27. Yu, J., et al., *Imaging Graphene Moire Superlattices via Scanning Kelvin Probe Microscopy.* Nano Lett, 2021. **21**(7): p. 3280-3286.

28. Shan, Y., et al., *Stacking symmetry governed second harmonic generation in graphene trilayers.* Science Advances, 2018. **4**(6): p. eaat0074.

29. Dean, J.J. and H.M. van Driel, *Second harmonic generation from graphene and graphitic films.* Applied Physics Letters, 2009. **95**(26).

30. Ohta, T., et al., *Interlayer Interaction and Electronic Screening in Multilayer Graphene Investigated with Angle-Resolved Photoemission Spectroscopy.* Physical Review Letters, 2007. **98**(20).

31. Taychatanapat, T., K. Watanabe, T. Taniguchi, and P. Jarillo-Herrero, *Quantum Hall effect and Landau-level crossing of Dirac fermions in trilayer graphene.* Nature Physics, 2011. **7**(8): p. 621-625.

32. Craciun, M.F., et al., *Trilayer graphene is a semimetal with a gate-tunable band overlap.* Nat Nanotechnol, 2009. **4**(6): p. 383-8.

33. Min, H. and A.H. MacDonald, *Origin of universal optical conductivity and optical stacking sequence identification in multilayer graphene.* Phys Rev Lett, 2009. **103**(6): p. 067402.

34. McEllistrim, A., A. Garcia-Ruiz, Z.A.H. Goodwin, and V.I. Fal'ko, *Spectroscopic signatures of tetralayer graphene polytypes.* Physical Review B, 2023. **107**(15).

35. Peres, N.M.R., *Colloquium: The transport properties of graphene: An introduction.* Reviews of Modern Physics, 2010. **82**(3): p. 2673-2700.

36. Falkovsky, L.A. and S.S. Pershoguba, *Optical far-infrared properties of a graphene monolayer and multilayer.* Physical Review B, 2007. **76**(15).

37. Mak, K.F., L. Ju, F. Wang, and T.F. Heinz, *Optical spectroscopy of graphene: From the far infrared to the ultraviolet.* Solid State Communications, 2012. **152**(15): p. 1341-1349.

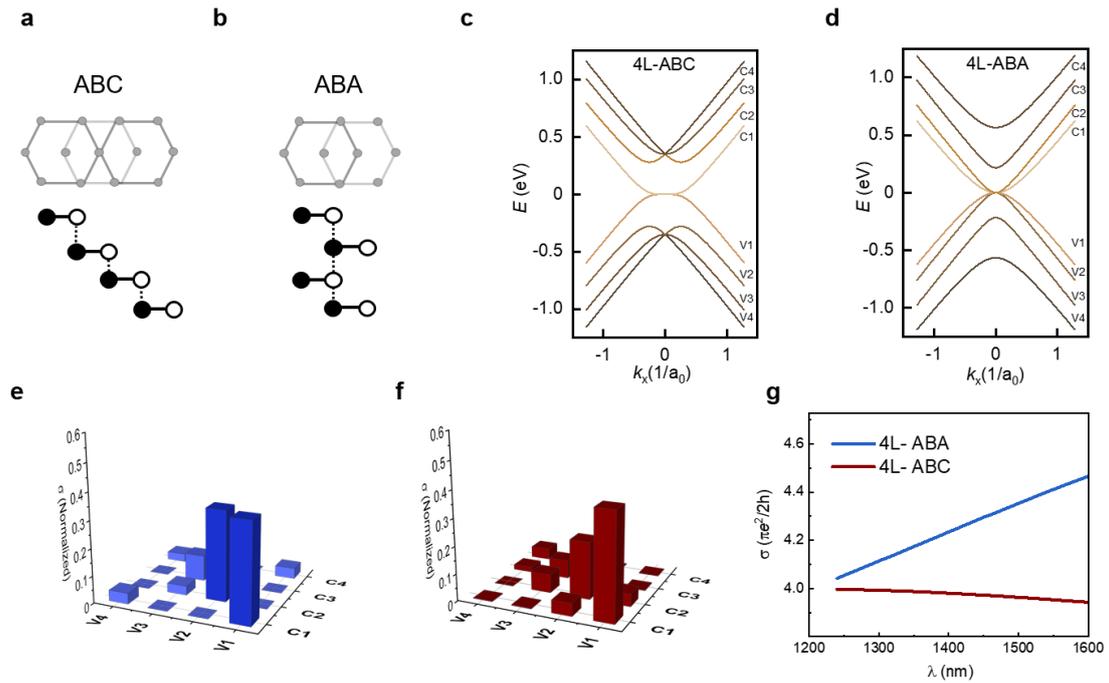

**Fig.1| The different optical reflectance in the infrared range of tetralayer ABC and ABA graphene**
**a,** Atomic structures of rhombohedral ABC stacked graphene. **b,** Atomic structures of Bernal ABA stacked graphene. **c,** Band structure of ABC tetralayer graphene. **d**, Band structure of ABA tetralayer graphene. **e**, Contributions of different band transitions to optical conductivity of ABC tetralayer graphene at 1600nm. **f,** Contributions of different band transitions to optical conductivity of ABA tetralayer graphene at 1600nm. **g,** Calculation results of the optical conductivity of ABA and ABC tetralayer graphene.

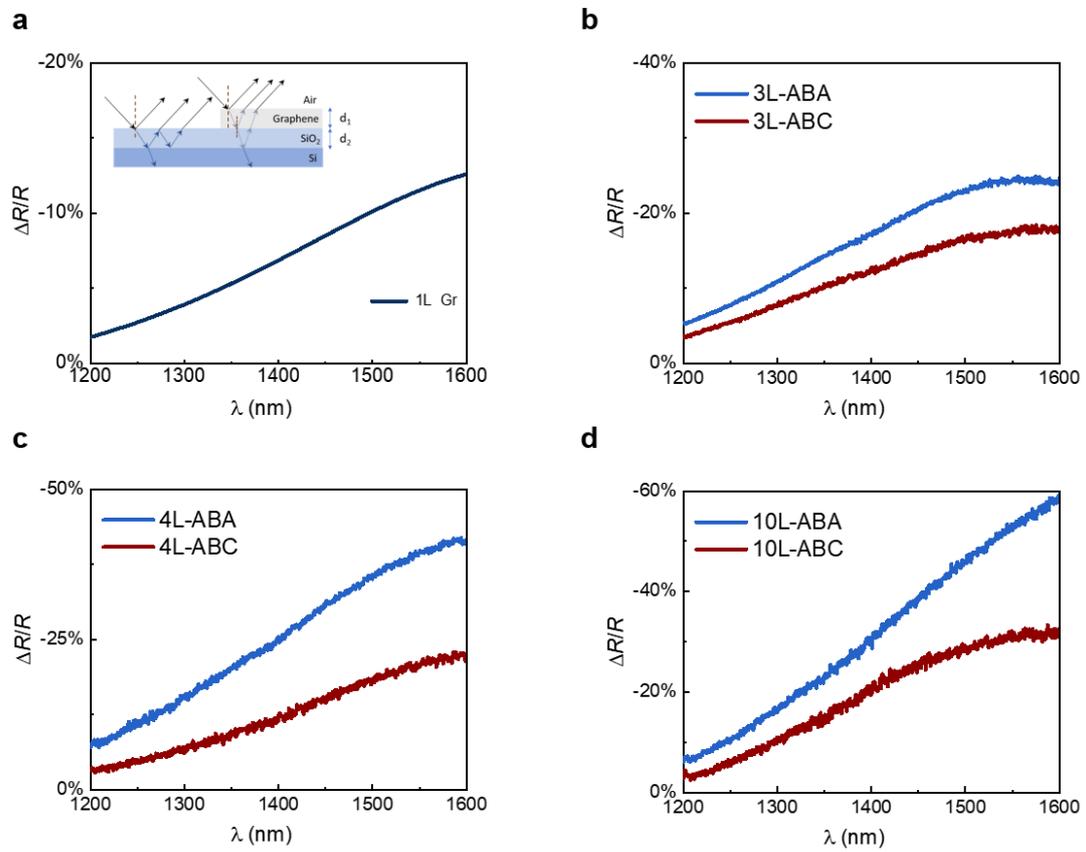

**Fig.2| The differential reflectance spectroscopy of ABA and ABC graphene. a,** Calculated contrast of monolayer graphene considering multi-reflection. Inset: Schematic of multi-reflection model. **b, c, d**, The differential reflectance spectroscopy of ABA and ABC trilayer graphene (**b**), tetralayer graphene (**c**) and decalayer graphene (**d**).

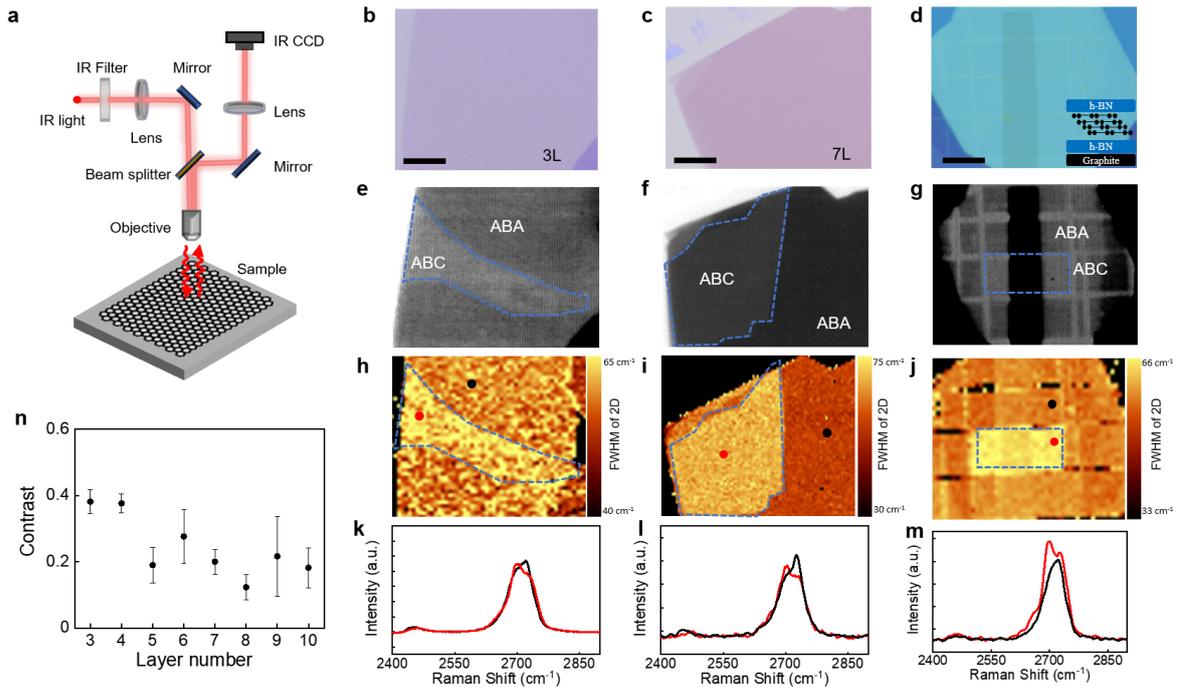

**Fig.3| Rapid infrared imaging system. a,** Schematic of infrared imaging system. **b,c,d,** Optical images of trilayer graphene (**b**) heptalayer graphene (**c**) and tetralayer graphene encapsulated by hBN with a bottom graphite gate (**d**). The scale bar is 10 μm. **e,f,g,** Infrared image of trilayer graphene (**e**), heptalayer graphene (**f**) and tetralayer graphene encapsulated by hBN (**g**). The brighter region shown in blue dashed line is the ABC stacked graphene. **h,i,j,** Raman mappings of trilayer graphene (**h**), heptalayer graphene (**i**) and tetralayer graphene encapsulated by hBN (**j**). The yellow region shown in blue dashed line corresponds to ABC stacking sequence, while the orange region corresponds to ABA stacking sequence. The mapping is color coded according to the FWHM of the 2D mode extracted from a single Lorentzian fit. **k,l,m,** Raman spectra (2D mode) of trilayer graphene (**k**), heptalayer graphene (**l**) and tetralayer graphene encapsulated by hBN (**m**). The red line represents the ABC stacking, and black line represents ABA stacking. **n,** Contrast difference between ABA and ABC graphene for different layers.

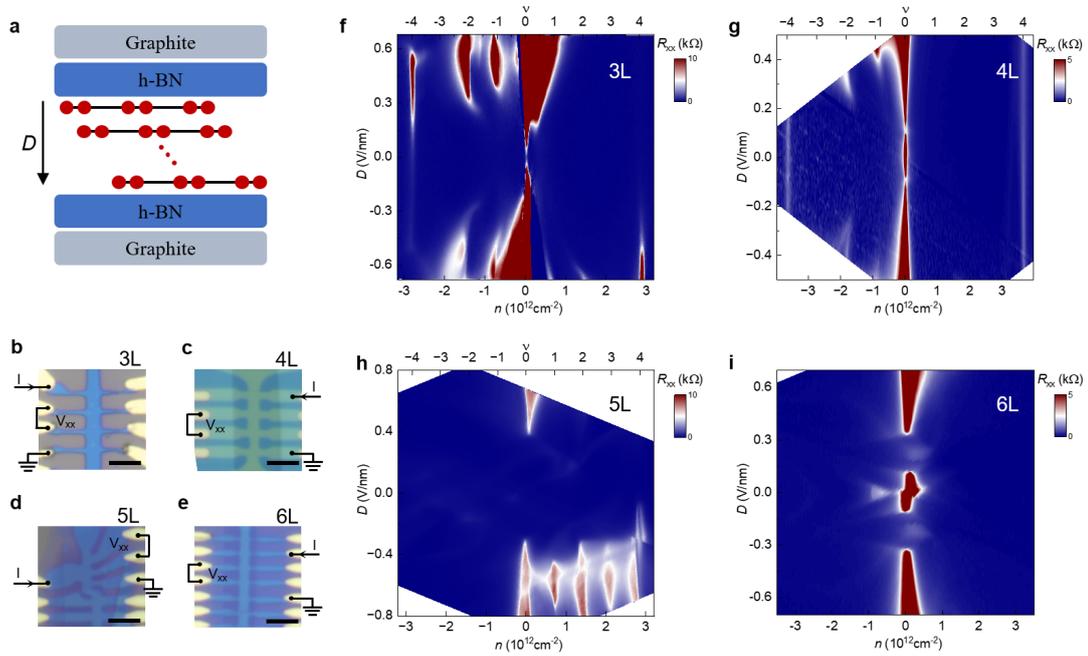

**Fig.4| Transport measurement of four different layer ABC stacked graphene. a,** Schematic of the dual-gated device. **b,c,d,e** Optical images of trilayer graphene (**b**) tetralayer graphene (**c**), pentalayer graphene (**d**) and hexalayer graphene (**e**). Scale bar is 5 μm. **f, g, h,** Color plot of $R_{xx}$ as a function of moiré filling $v$ and displacement field $D$ for trilayer graphene (**f**), heptalayer graphene (**g**) and pentalayer graphene (**h**). **I,** Color plot of $R_{xx}$ as a function of carrier density $n$ and displacement field $D$ for hexalayer graphene.

Supplementary Materials for

# Rapid infrared imaging for rhombohedral graphene

Zuo Feng[1,2†], Wenxuan Wang[2†], Yilong You[1,2†], Yifei Chen[3], Kenji Watanabe[4], Takashi Taniguchi[5], Chang Liu[1,2]*, Kaihui Liu[1,2]* and Xiaobo Lu[2,6]*

[1]State Key Laboratory for Mesoscopic Physics, Frontiers Science Centre for Nano-optoelectronics, School of Physics, Peking University, 100871, Beijing, China
[2]International Center for Quantum Materials, School of Physics, Peking University, Beijing 100871, China
[3]International School of Beijing, Beijing, 101318
[4]Research Center for Electronic and Optical Materials, National Institute of Material Sciences, 1-1 Namiki, Tsukuba 305-0044, Japan
[5]Research Center for Materials Nanoarchitectonics, National Institute of Material Sciences, 1-1 Namiki, Tsukuba 305-0044, Japan
[6]Collaborative Innovation Center of Quantum Matter, Beijing 100871, China

[†]Z.F., W.W., and Y.L. contributed equally to this work.
*E-mail: xiaobolu@pku.edu.cn; liu.chang@pku.edu.cn; khliu@pku.edu.cn

**This PDF file includes:**

**Supplementary Note 1,2,3**

**Supplementary Note 1: Calculation of Band Structure**

We use the tight-binding model to calculate the band structure of multilayer graphene [1]. Here we take the basis as $|A_1\rangle, |B_1\rangle; |A_2\rangle, |B_2\rangle; ...; |A_N\rangle, |B_N\rangle$. In order to capture the main features of our experiment data, we only consider the nearest intra-layer hopping parameter $\gamma_0$ and inter-layer hopping parameter $\gamma_1$. Then the Hamiltonian around the K point becomes

$$\mathcal{H} = \begin{pmatrix} H_0 & V & & \\ V^\dagger & H_0 & V^\dagger & \\ & V & H_0 & V \\ & & \ddots & \ddots & \ddots \end{pmatrix}, \qquad (1)$$

with

$$H_0 = \begin{pmatrix} 0 & vp_- \\ vp_+ & 0 \end{pmatrix}, V = \begin{pmatrix} 0 & 0 \\ \gamma_1 & 0 \end{pmatrix}, \qquad (2)$$

where $p_\pm = p_x \pm ip_y$, with $\boldsymbol{p} = -i\hbar\boldsymbol{\nabla}$, and the velocity of monolayer graphene is $v$. According to former researches, $v$ is related to the band parameter through the equation $v = \sqrt{3}a\gamma_0/2\hbar$. $H_0$ describes the intra-layer hopping, while $V$ depicites the inter-layer hopping. Here we use $a = 0.246nm$, $\gamma_0 = 3.16eV$ and $\gamma_1 = 0.37eV$ [2].

Especially in 4-layer graphene as is mentioned in the article, the Hamiltonian for AB stacked graphene is

$$\mathcal{H} = \begin{pmatrix} H_0 & V & & \\ V^\dagger & H_0 & V^\dagger & \\ & V & H_0 & V \\ & & V^\dagger & H_0 \end{pmatrix}, \qquad (3)$$

and the Hamiltonian for ABC-stacked graphene is

$$\mathcal{H} = \begin{pmatrix} H_0 & V & & \\ V^\dagger & H_0 & V & \\ & V^\dagger & H_0 & V \\ & & V^\dagger & H_0 \end{pmatrix}, \qquad (4)$$

**Supplementary Note 2: Calculation of optical conductivity**

We employ the Kubo formula to calculate the optical conductivity:

$$\sigma(\omega)_{i,j} = \frac{e^2 \hbar}{iS} \sum_{b_1,b_2,k} \frac{n_f(e^k_{b_1}) - n_f(e^k_{b_2})}{e^k_{b_1} - e^k_{b_2}} \frac{\langle e^k_{b_1}|v_i|e^k_{b_2}\rangle \langle e^k_{b_2}|v_j|e^k_{b_1}\rangle}{\hbar\omega + e^k_{b_1} - e^k_{b_2} + i\eta}, \quad (5)$$

where $e^k_b$ are the eigenenergies of the Hamiltonian in band b at momentum point k and $|e^k_b\rangle$ are the corresponding eigenvectors. $n_f$ refers to the Fermi–Dirac distribution [3].

In our calculation, we only consider the x-direction optical conductivity $v_x = -(i/\hbar)[x, \mathcal{H}] = \partial \mathcal{H} / \partial p_x$. While obviously $\partial V / \partial p_x = 0$, which means $v_x$ is

$$v_x = \begin{pmatrix} \partial H_0 / \partial p_x & & & \\ & \partial H_0 / \partial p_x & & \\ & & \partial H_0 / \partial p_x & \\ & & & \partial H_0 / \partial p_x \end{pmatrix}, \quad (6)$$

Here we take $\eta = 20 meV$ and $T = 300K$ into our calculation.

**Supplementary Note 3: Calculation of multi-refection process**

To explain the contrast change of graphene on SiO$_2$/Si substrate. We use the Fresnel's equations. The multireflection model is composed of graphene, SiO$_2$ and Silicon layers. The light is incident from air onto the three layers, the reflected light intensity can be calculated using the following equation:

$$R_{gr} = \left| \frac{r_1 + r_2 e^{-2i\beta_1} + r_3 e^{-2i(\beta_1+\beta_2)} + r_1 r_2 r_3 e^{-2i\beta_2}}{1 + r_1 r_2 e^{-2i\beta_1} + r_1 r_3 e^{-2i(\beta_1+\beta_2)} + r_2 r_3 e^{-2i\beta_2}} \right|^2$$

$$R_{sub} = \left| \frac{(r_2 + r_3 e^{-2i\beta_2})}{1 + r_2 r_3 e^{-2i\beta_2}} \right|^2$$

$$\frac{\Delta R}{R} = \frac{R_{gr} - R_{sub}}{R_{sub}}$$

Where

$$r_2 = \frac{n_{gr} - n_3}{n_{gr} + n_3}, r_2 = \frac{n_{gr} - n_3}{n_{gr} + n_3}, r_3 = \frac{n_2 - n_3}{n_2 + n_3}$$

$$\beta_1 = 2\pi n_{gr} d_1 / \lambda, \beta_2 = 2\pi n_2 d_2 / \lambda.$$

Considering the refractive index can be described by the following equation:

$$\sigma = i\omega(n_{gr}^2 - \varepsilon_0),$$

where σ is the optical conductivity of graphene, ω is the frequency of the incident light, ε$_0$ is the dielectric constant of vacuum.

The reflected light intensity depended on optical conductivity can be described by the following equation:

$$R_{gr}(\sigma) = \left| \frac{\frac{n_0 - \sqrt{\frac{\sigma}{i\omega} + \varepsilon_0}}{n_0 + \sqrt{\frac{\sigma}{i\omega} + \varepsilon_0}} \cdot \frac{\sqrt{\frac{\sigma}{i\omega} + \varepsilon_0} - n_2}{\sqrt{\frac{\sigma}{i\omega} + \varepsilon_0} + n_2} e^{-2i\beta_1} + \frac{n_2 - n_3}{n_2 + n_3} e^{-2i(\beta_1+\beta_2)} + \frac{n_0 - \sqrt{\frac{\sigma}{i\omega} + \varepsilon_0}}{n_0 + \sqrt{\frac{\sigma}{i\omega} + \varepsilon_0}} \cdot \frac{\sqrt{\frac{\sigma}{i\omega} + \varepsilon_0} - n_2}{\sqrt{\frac{\sigma}{i\omega} + \epsilon} + n_2} \cdot \frac{n_2 - n_3}{n_2 + n_3} e^{-2i\beta_2}}{1 + \frac{n_0 - \sqrt{\frac{\sigma}{i\omega} + \varepsilon_0}}{n_0 + \sqrt{\frac{\sigma}{i\omega} + \varepsilon_0}} \cdot \frac{\sqrt{\frac{\sigma}{i\omega} + \varepsilon_0} - n_2}{\sqrt{\frac{\sigma}{i\omega} + \varepsilon_0} + n_2} e^{-2i\beta_1} + \frac{n_0 - \sqrt{\frac{\sigma}{i\omega} + \varepsilon_0}}{n_0 + \sqrt{\frac{\sigma}{i\omega} + \varepsilon_0}} \cdot \frac{n_2 - n_3}{n_2 + n_3} e^{-2i(\beta_1+\beta_2)} + \frac{\sqrt{\frac{\sigma}{i\omega} + \varepsilon_0} - n_2}{\sqrt{\frac{\sigma}{i\omega} + \varepsilon_0} + n_2} \cdot \frac{n_2 - n_3}{n_2 + n_3} e^{-2i\beta_2}} \right|^2,$$

$$R_{sub} = \left| \frac{\frac{\sqrt{\frac{\sigma}{i\omega} + \varepsilon_0} - n_2}{\sqrt{\frac{\sigma}{i\omega} + \varepsilon_0} + n_2} + \frac{n_2 - n_3}{n_2 + n_3} e^{-2i\beta_2}}{1 + \frac{n_0 - \sqrt{\frac{\sigma}{i\omega} + \varepsilon_0}}{n_0 + \sqrt{\frac{\sigma}{i\omega} + \varepsilon_0}} \cdot \frac{n_2 - n_3}{n_2 + n_3} e^{-2i\beta_2}} \right|^2,$$

## References

[1] Koshino, M. and Ando, T., 2008. Magneto-optical properties of multilayer graphene. Physical Review B, 77(11), p.115313.

[2] Dresselhaus, M.S. and Dresselhaus, G., 1981. Intercalation compounds of graphite. Advances in Physics, 30(2), pp.139-326.

[3] Kubo, Ryogo., 1957. Statistical-mechanical theory of irreversible processes. I. General theory and simple applications to magnetic and conduction problems. Journal of the physical society of Japan 12(6), pp.570-586.